\documentclass[12pt]{article}
\usepackage{epsf}
\catcode`@=11
\def\citer{\@ifnextchar
[{\@tempswatrue\@citexr}{\@tempswafalse\@citexr[]}}

\def\@citexr[#1]#2{\if@filesw\immediate\write\@auxout{\string\citation{#2}}\fi
  \def\@citea{}\@cite{\@for\@citeb:=#2\do
    {\@citea\def\@citea{--\penalty\@m}\@ifundefined
       {b@\@citeb}{{\bf ?}\@warning
       {Citation `\@citeb' on page \thepage \space undefined}}%
\hbox{\csname b@\@citeb\endcsname}}}{#1}}
\catcode`@=12
\def\Journal#1#2#3#4{{#1} {\bf #2}, #3 (#4)}
\def\NPB{{\em Nucl.\ Phys.}\ B}
\def\NPPS{{\em Nucl.\ Phys.\ Proc.\ Suppl.}\ }
\def\PLB{{\em Phys.\ Lett.}\ B}
\def\PL{{\em Phys.\ Lett.}\ }
\def\PRL{{\em Phys.\ Rev.\ Lett.}\ }
\def\PRD{{\em Phys.\ Rev.}\ D}

\def\JPG{{\em J.\ Phys.}\ G}
\newcommand{\md}{\mbox{d}}
\newcommand{\beq}{\begin{equation}}
\newcommand{\eeq}{\end{equation}}
\newcommand{\bea}{\begin{eqnarray}}
\newcommand{\eea}{\end{eqnarray}}

\newcommand{\alps}{\mbox{$\alpha_{\mbox{\scriptsize s}}$}}
\newcommand{\alpsq}{\mbox{$\alpha_{\mbox{\scriptsize s}}^{2}$}}
\newcommand{\alptr}{\mbox{$\alpha_{\mbox{\scriptsize s}}^{3}$}}
\newcommand{\alpsf}{\mbox{$\alpha_{\mbox{\scriptsize s}}^{4}$}}
\def\sz{\scriptsize}
\def\simgt{\rlap{\lower 3.5 pt \hbox{$\mathchar \sim$}} 
                                \raise 1pt \hbox {$>$}}
\def\simlt{\rlap{\lower 3.5 pt \hbox{$\mathchar \sim$}} 
                                \raise 1pt \hbox {$<$}}
\begin{document}

\begin{titlepage}

\begin{flushright}
CERN-TH/99-105\\
April 1999
\end{flushright}

\vspace*{1cm}

\begin{center} {\Large QCD Corrections to Radiative $\Upsilon$ Decays}

\vspace*{1cm}

{\sc Michael Kr\"amer}

\vspace*{2mm}

{\it Theoretical Physics Division, CERN, CH-1211 Geneva 23, Switzerland}

\end{center}

\vspace*{1cm}

\begin{abstract} 
We have calculated the next-to-leading order perturbative QCD
corrections to the photon energy spectrum in radiative $\Upsilon$
decays. The higher-order corrections significantly modify the shape of
the spectrum, in particular at large photon energies, and thereby
reduce the discrepancy between experimental data and previous
leading-order predictions. The next-to-leading order calculation of
the photon energy spectrum allows a more reliable determination of the
strong coupling constant from radiative $\Upsilon$ decays by
restricting the analysis to the region of the energy spectrum that can
be described by NLO perturbation theory.
\end{abstract}

\vfill

\begin{flushleft}
CERN-TH/99-105\\
April 1999
\end{flushleft}

\end{titlepage}

{\bf 1. Introduction.}
The calculations of heavy quarkonium decay rates are among the
earliest applications of perturbative QCD~\cite{AP-75}; they have been
used to extract information on the QCD coupling at scales of the order
of the heavy-quark mass. A consistent and rigorous framework for
treating inclusive quarkonium decays has recently been developed,
superseding the earlier non-relativistic potential model.  The
factorization approach \cite{BBL-95} is based on the use of
non-relativistic QCD (NRQCD) to separate the short-distance physics of
heavy-quark annihilation from the long-distance physics of bound-state
dynamics. The quarkonium decay rate can be expressed as a sum of
terms, each of which factors into a short-distance coefficient and a
long-distance matrix element. The short-distance coefficients are
determined by the annihilation cross section for an on-shell
$Q\overline{Q}[n]$ pair in a colour, spin and angular-momentum
configuration $n$, and can be calculated perturbatively in the strong
coupling $\alps(m_Q)$. The probability to find a $Q\overline{Q}$ pair
in the quarkonium at the same point and in the configuration $n$ is
parametrized by a non-perturbative matrix element $\langle {\cal
O}_{n}^{\Upsilon}\rangle $, which is subject to the power counting
rules of NRQCD and which can be evaluated numerically using NRQCD
lattice simulations \cite{BSK-96}. It is thus, in principle, possible
to calculate the annihilation decay rates of heavy quarkonium from
first principles, the only inputs being the heavy-quark mass $m_Q$ and
the QCD coupling $\alps(m_Q)$. Fixing the heavy-quark mass from, for
example, sum rule calculations for quarkonia, or considering ratios of
different decay channels where the quark-mass dependence and the
long-distance factor cancel, the analysis of quarkonium decay rates
can provide information on the QCD coupling at the heavy-quark mass
scale.

In this letter, we focus on the reaction $\Upsilon \to \gamma + X$,
which is used for $\alps$ determinations from the quarkonium system
and which provides detailed information on the decay dynamics through
the analysis of the photon energy spectrum. We present the first
calculation of the next-to-leading order perturbative QCD corrections
to the photon spectrum. Besides being an interesting problem of
perturbative QCD per se, the inclusion of higher-order corrections
allows a NLO determination of the strong coupling constant, avoiding
those regions of the energy spectrum that are affected by
non-perturbative contributions.\footnote{In the absence of a NLO
prediction for the photon spectrum, model approaches have been used to
extrapolate the experimental data into the region of low photon
energies to obtain the total decay rate, introducing uncontrolled
theoretical uncertainties. The model dependence of the corresponding
NLO determination of $\alps$ can be avoided by restricting the
analysis to the region of the energy spectrum that is described by NLO
perturbation theory.}

\vspace*{3mm}

\begin{figure}[htb]
   \vspace*{-3.5cm}
   \epsfysize=25cm
   \epsfxsize=20cm
   \centerline{\epsffile{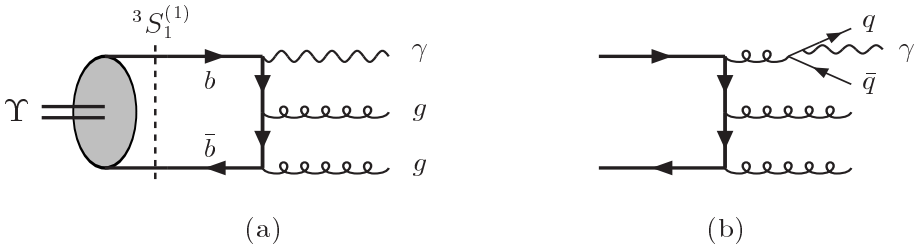}}
   \vspace*{-18cm}
\caption[dummy]{\small \label{fig:feyn-lo} Generic leading-order Feynman 
diagrams for radiative $\Upsilon$ decay: (a) direct contribution, (b)
fragmentation contribution.}
\end{figure}

{\bf 2. The leading-order calculation.}
At leading order in the non-relativistic velocity expansion in $v^2$,
radiative $\Upsilon$ decays proceed through the annihilation of a
colour-singlet $n={}^{3}S_1^{(1)}$ $b\bar{b}$ pair (in spectroscopic
notation).\footnote{At leading order in the velocity expansion, the
NRQCD description of $S$-wave quarkonium decays is equivalent to the
non-relativistic potential model, where the non-perturbative dynamics
is described by a single long-distance factor related to the
bound-state's wave function at the origin.  In the case of $P$-wave
quarkonia, the potential-model calculations at ${\cal O}(\alptr)$
encounter infrared divergences, which cannot be factored into a single
non-perturbative quantity. Within the NRQCD approach, this problem
finds its natural solution since the infrared singularities are
factored into a long-distance matrix element related to higher
Fock-state components of the quarkonium wave function.}  The photon
energy spectrum receives contributions from two different
mechanisms. At leading oder, one has
\beq 
 \frac{\md\Gamma^{\mbox{\sz LO}}(\Upsilon\to \gamma X)}{\md x_\gamma} 
= \frac{\md\Gamma_{\mbox{\sz dir}}^{\mbox{\sz LO}}}{\md x_\gamma} 
+ \int_{x_\gamma}^1\,\frac{\md{z}}{z}\,C_{g}(z)\, D_g^{\gamma}
\left(\frac{x_\gamma}{z}\right) ,
\eeq
where $0\leq x_\gamma = E_\gamma/m_b \leq 1$. The first term on the
right-hand side denotes the direct cross section, where the photon is
radiated off a heavy quark, Fig.~\ref{fig:feyn-lo}(a), while the
second term represents the contribution from gluon fragmentation into
a photon, Fig.~\ref{fig:feyn-lo}(b).
%-------------------------------------------------------------------------
%-------------------------------------------------------------------------
\noindent The fragmentation contribution, although of ${\cal
O}(\alpha\alpsf)$ in the coupling constant, is enhanced by a double
logarithmic singularity $\sim \ln^2(m_b^2/\Lambda^2)\sim 1/\alpsq$
arising from the phase-space region where both the
light-quark--antiquark splitting as well as the photon emission become
collinear \cite{CH-95}. The fragmentation function $D_g^{\gamma}$ is
sensitive to non-perturbative effects and has to be extracted from
experiment. A comparison between the direct component and the
fragmentation contribution to the photon spectrum, performed at
leading order, reveals that the fragmentation process is important in
the low-$x_\gamma$ region, but suppressed with respect to the direct
cross section for $x_\gamma\;\simgt\;0.3$
\citer{CH-95,MP-98}.\footnote{At next-to-leading order, the fragmentation 
contribution involves also quark fragmentation, which is much harder
than gluon fragmentation and may be important at larger values of
$x_\gamma$.}  It is, however, important to point out that the total
decay rate, including fragmentation processes,
$\Gamma(\Upsilon\to\gamma X) = \int_0^1 \, \md x_\gamma \, \md \Gamma
/ \md x_\gamma $, is not an infrared-safe observable. For $x_\gamma\to
0$, the fragmentation contribution rises like $1/x_\gamma$,
characteristic of the soft bremsstrahlung spectrum, and cannot be
integrated down to $x_\gamma =0$. Therefore, in perturbation theory,
only the photon energy spectrum can be calculated for $x_\gamma \neq
0$, after collinear singularities have been factorized into the
fragmentation functions.

In the following, we will focus on the direct contribution of the
energy spectrum, which dominates in the region $x_\gamma
\;\simgt\;0.3$ where accurate experimental data are available.  At
leading order, one finds \cite{BGHC-78}
\begin{eqnarray}\label{eq:spec-lo}
\hspace*{-1cm}\frac{1}{\Gamma_{\mbox{\sz dir}}^{\mbox{\sz LO}}}
\frac{\md\Gamma_{\mbox{\sz dir}}^{\mbox{\sz LO}}(\Upsilon\to 
\gamma X)}{\md x_\gamma} &=& 
\frac{2}{(\pi^2-9)}\left(
2\,\frac{1-x_\gamma}{x_\gamma^2}\,\ln(1-x_\gamma)
 -2\,\frac{(1-x_\gamma)^2}{(2-x_\gamma)^3}\,
\ln(1-x_\gamma) 
\right.\nonumber\\
\nonumber\\&&\hspace*{2cm}
\left.
+\frac{2-x_\gamma}{x_\gamma}
+\frac{x_\gamma (1-x_\gamma)}{(2-x_\gamma)^2}\right).
\end{eqnarray}
To very good accuracy, Eq.~(\ref{eq:spec-lo}) can be approximated by a
linear spectrum: $1/\Gamma_{\mbox{\sz dir}}^{\mbox{\sz LO}}$ $\md
\Gamma_{\mbox{\sz dir}}^{\mbox{\sz LO}}/\md x_\gamma$ $\approx$ 
$2 x_\gamma$. The corresponding leading-order decay rate is given by
\cite{CHAN-75}
\begin{equation}\label{eq:rate-lo}
\Gamma_{\mbox{\sz dir}}^{\mbox{\sz LO}}(\Upsilon\to \gamma X) = 
\frac{16}{27}\,\alpha\,\alpsq\, e_b^2
\,(\pi^2-9)\,\frac{\langle {\cal O}^{\Upsilon}_{1}({}^3S_1) 
\rangle}{m_b^2} ,
\end{equation}
where $e_b$ is the charge of the $b$ quark. Up to corrections of
${\cal O}(v^4)$, the non-perturbative NRQCD matrix element is related
to the $\Upsilon$ wave function at the origin through $\langle {\cal
O}^{\Upsilon}_{1}({}^3S_1) \rangle$ $\approx$ $2 N_C \,|\,
\varphi(0)\,|^2$, with $N_C=3$ the number of colours.

The linear rise with $x_\gamma$ predicted by LO perturbative QCD,
Eq.~(\ref{eq:spec-lo}), is not supported by the experimental data
\cite{data}. In particular for $x_\gamma\;\simgt\; 0.7$ the
discrepancy is significant.  The leading-order estimate of the photon
energy spectrum demands several theoretical refinements: the inclusion
of (i) higher-order perturbative QCD corrections; (ii) relativistic
corrections due to the motion of the $b$ quarks in the $\Upsilon$
bound state; and (iii) hadronization effects that may become
important close to the phase-space boundary. The relativistic
corrections to the photon energy spectrum, including colour-octet
contributions, have been analysed in \cite{rel,MP-98}. Although
potentially sizeable, they are concentrated in the upper and lower
end-point region and do not significantly modify the shape of the
spectrum in the intermediate energy range $0.4 \;\simlt \; x_\gamma
\;\simlt\; 0.9$.  Hadronization effects have been studied in the
context of different models \cite{hadron} and may become important
when $x_\gamma \to 1$. However, these non-perturbative contributions
are not expected to modify the energy distribution significantly in
the intermediate region $x_\gamma\sim 0.7$. The calculation of the
next-to-leading order perturbative QCD corrections to the photon
spectrum is the subject of this letter.

\vspace*{3mm}

\begin{figure}[htb]
   \vspace*{-4.5cm}
   \epsfysize=25.75cm
   \epsfxsize=20cm
   \centerline{\epsffile{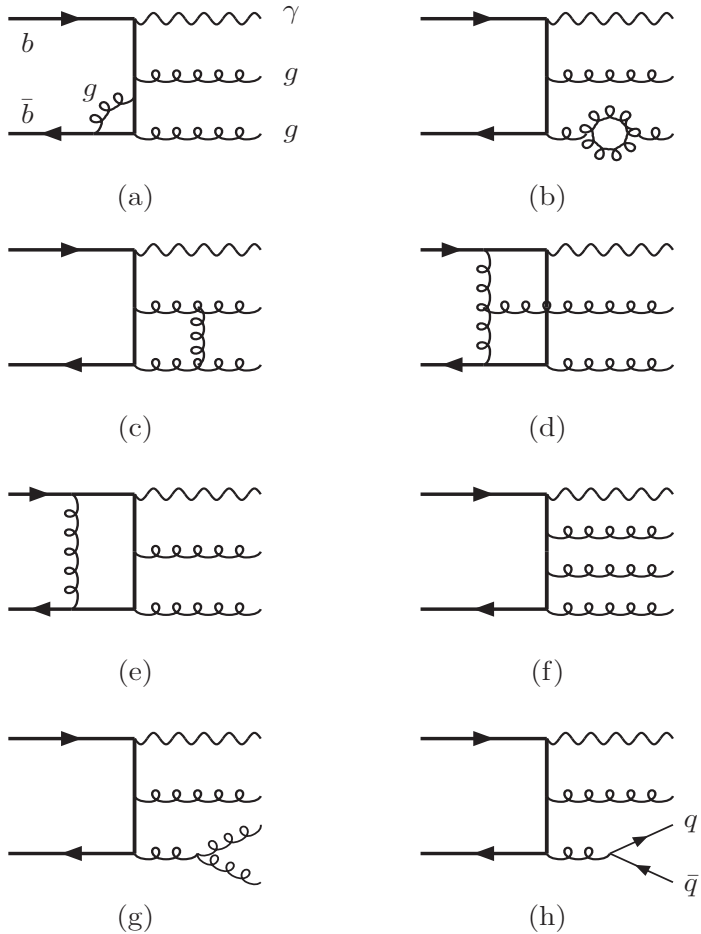}}
   \vspace*{-9cm}
\caption[dummy]{\small \label{fig:feyn-nlo} Generic next-to-leading order 
 Feynman diagrams for direct radiative $\Upsilon$ decay.}
\vspace*{-2mm}
\end{figure}

{\bf 3. Next-to-leading order QCD corrections.}
Generic diagrams that build up the decay rate in ${\cal
O}(\alpha\alptr)$ are drawn in Fig.~\ref{fig:feyn-nlo}.
%-------------------------------------------------------------------------
%-------------------------------------------------------------------------
The evaluation of the next-to-leading order corrections involves two
parts, the virtual corrections, generated by virtual particle
exchange, and the real corrections, which originate from real gluon
radiation as well as from gluon splitting into light-quark--antiquark
pairs \cite{KZSZ-95}.

The virtual amplitude consist of 105 diagrams, including self-energy
and vertex corrections for photon and gluons (2a,b) as well as a large
number of box diagrams (2c--e). The evaluation of these amplitudes has
been performed in the Feynman gauge. The ultraviolet and infrared
singularities have been extracted using dimensional regularization,
leading to poles in $\epsilon = (n-4)/2$. The Feynman integrals
containing loop momenta in the numerator have been reduced to a set of
scalar integrals using an adapted version of the reduction program
outlined in \cite{PV-79}.  This program has been extended to treat
$n$-dimensional tensor integrals with linear-dependent propagators to
account for the special kinematical situation in the non-relativistic
approximation to $\Upsilon$ decays. Using the Feynman parameter
technique, we have obtained analytical results for the scalar
integrals; these will be presented in a forthcoming publication. The
masses of the $n_{\mbox{\sz lf}} = 4$ light quarks have been
neglected, while the mass parameter of the $b$ quark has been defined
on-shell. We have carried out the renormalization of the QCD coupling
in the $\overline{\rm MS}$ scheme, including $n_{\mbox{\sz lf}} = 4$
light flavours in the corresponding $\beta$ function.  The bottom
quark has been decoupled from the evolution of $\alps$ by subtracting
its contribution at vanishing momentum transfer \cite{alphas}.  The
exchange of Coulombic gluons in diagram (2e) leads to a singularity
$\sim \pi^2/2\beta_R$, which can be isolated by introducing a small
relative quark velocity $\beta_R$. The Coulomb-singular part can be
associated with the interquark potential of the bound state and has to
be factored into the non-perturbative NRQCD matrix element. Only the
exchange of transverse gluons contributes to the next-to-leading order
expression for the short-distance annihilation rate.

The real corrections are generated by gluon radiation off quarks and
gluons (2f,g) and by gluon splitting into light-quark--antiquark pairs
(2h).  The 54 diagrams have been calculated in the Feynman gauge. For
the sum of the gluon polarization we have used $ \sum
\varepsilon_{\mu} \varepsilon_{\nu} = -g_{\mu\nu}$ and the unphysical
longitudinal gluon polarizations have been removed by adding ghost
contributions. The real emission cross section contains singularities
in the limit where the gluon splitting into gluon or
light-quark--antiquark pairs becomes soft or collinear. We have used
the subtraction method \cite{ERT-81} as formulated in \cite{CS-97} to
extract the singular parts of the real cross section in $n$
dimensions.  The method of \cite{CS-97} is based on the fact that the
soft and collinear limits of a real emission matrix element can be
expressed in terms of a convolution of process-independent splitting
functions and the leading-order matrix element. Subtracting from the
real matrix element counterterms constructed accordingly, the infrared
and collinear singularities cancel and the phase-space integration can
be performed numerically in four dimensions. The counterterms can be
integrated analytically in $n$ dimensions over the phase space of the
extra emitted parton, leading to poles in $\epsilon = (n-4)/2$. When
the integral of the subtraction counterterm and the virtual cross
section are combined, the infrared and collinear singularities
cancel. The analytical result for the matrix element squared has been
implemented in a Monte Carlo integration program, so that not only the
inclusive decay rate but any distribution can be generated at
next-to-leading order.

As mentioned above, owing to the presence of fragmentation processes
at small photon energies, the total decay rate is not an infrared-safe
observable. We will nevertheless present a next-to-leading order
result for the perturbatively calculable direct contribution to the
total decay rate, in order to compare our calculation with a previous
NLO estimate of that quantity. We find
\beq\label{eq:rate-nlo2}
\Gamma_{\mbox{\sz dir}}^{\mbox{\sz NLO}}(\Upsilon\to \gamma X) =
\Gamma_{\mbox{\sz dir}}^{\mbox{\sz LO}}\,
\left[1-\frac{\alps(\mu)}{\pi}\left(1.71
+\beta_0(n_{\mbox{\sz lf}})
\ln\frac{2m_b}{\mu}\right)\right] ,
\eeq
where $\beta_0(n_{\mbox{\sz lf}}) = (11N_C-2n_{\mbox{\sz lf}})/3$. The
theoretical uncertainty due to the numerical phase-space integration
has been reduced to $\simlt\;0.5\%$. The new, more accurate result for
the total rate is consistent with the previous calculation
\cite{ML-81}, which yielded 
$ \Gamma_{\mbox{\sz dir}}^{\mbox{\sz NLO}} / 
  \Gamma_{\mbox{\sz dir}}^{\mbox{\sz LO}} 
  = (1- \alpha_s (1.67\pm 0.36)/\pi) $ 
for $\mu = 2m_b$, with a considerable theoretical uncertainty of
$\sim\pm 20\%$ coming from the numerical evaluation of the loop
integrals.\footnote{A prescription for fixing the renormalization
scale $\mu$ has been given in \cite{BLM-83}. The analysis implies that
it is advantageous to consider ratios like $\Gamma(\Upsilon\to \gamma
g g)/\Gamma(\Upsilon\to ggg)$ for $\alps$ extractions.}

\begin{figure}[htb]
   \vspace*{5mm}
   \epsfysize=10cm
   \epsfxsize=12cm
   \centerline{\epsffile{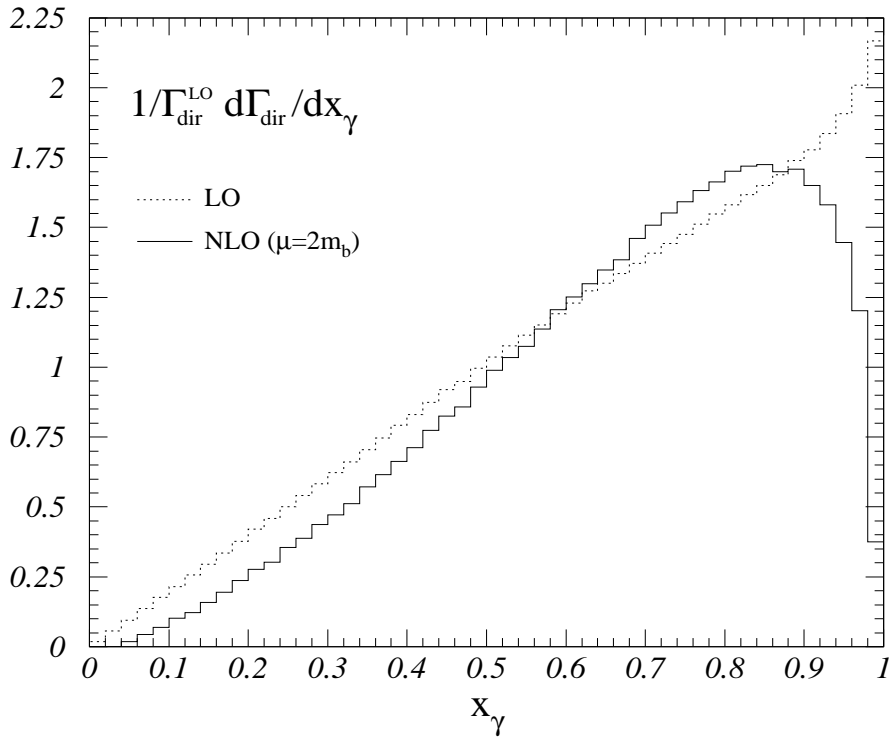}}
\caption[dummy]{\small \label{fig:spec-nlo} Photon energy spectrum 
for direct radiative $\Upsilon$ decays in leading and next-to-leading
order ($\alps = 0.2$).}
\end{figure}

The central part of this work is the calculation of the photon energy
spectrum at next-to-leading order. The result is presented in
Fig.~\ref{fig:spec-nlo}. The QCD corrections significantly flatten and
%-------------------------------------------------------------------------
%-------------------------------------------------------------------------
deplete the spectrum for $x_\gamma\;\simgt\;0.75$.  The shape of the
NLO spectrum is in qualitative agreement with the experimental data
and indicates that the discrepancy between theory and experiment at
large photon energies can be reduced by the inclusion of higher-order
QCD corrections. A proper comparison with the recent measurements
requires the inclusion of experimental aspects, such as energy
resolution and efficiency, and will be presented in a forthcoming
publication.

Finally, let us briefly comment on the general structure of the
higher-order corrections near the upper end-point of the photon energy
spectrum. It has been argued that potentially large logarithms
$\ln(1-x_\gamma)$ associated with the imperfect cancellation between
real and virtual emission of soft gluons as $x_\gamma \to 1$ may
contribute to all orders in perturbation theory \cite{PHOT-85}. The
resummation of these soft-gluon effects may then give rise to a
Sudakov suppression $\sim 1/\exp(\alps\ln^2(1-x_\gamma))$ near the
end-point of the photon energy spectrum. In contrast to the previous
claim \cite{PHOT-85}, a recent analysis \cite{CHM-00} finds that the
logarithms of $(1-x_\gamma)$ cancel at each order in the perturbative
expansion, as long as the $\Upsilon$ decay proceeds through the
annihilation of a colour-singlet $b\bar{b}$ pair. While the numerical
NLO calculation of the spectrum as presented in
Fig.~\ref{fig:spec-nlo} is not suited to definitely resolving that
issue, it may be possible to derive, from the full NLO result, a
simple analytic expression of the NLO cross section near the end-point
region. We leave such a calculation for a forthcoming publication.

\newpage

In conclusion, the photon spectrum in radiative $\Upsilon$ decays is a
very interesting laboratory to study the effects of higher-order
perturbative QCD corrections. The next-to-leading order calculation
presented here shows that the intermediate region of the energy
spectrum can be adequately accounted for by perturbative QCD.
Fragmentation processes, relativistic corrections and hadronization
effects have to be included to further improve the theoretical
description of the spectrum in the end-point regions.  In this
respect, the NLO calculation is vital to disentangle perturbative
effects from non-perturbative physics and relativistic corrections,
and thus test the hadronization models that have been suggested in the
literature. Finally, a more reliable NLO extraction of $\alps$ from
radiative $\Upsilon$ decays is possible by restricting the analysis to
the perturbatively calculable part of the energy spectrum.

\vspace*{3mm}

\small
\noindent
{Acknowledgements}

\noindent
I would like to thank J\"urgen Steegborn for his collaboration during
earlier stages of this work and Johann K\"uhn for his encouragement
and critical reading of the manuscript. Special thanks go to Stefan
Dittmaier for an independent evaluation of the scalar five-point
integral.  I have benefited from discussions with Stefano Catani,
Francesco Hautmann and Fabio Maltoni.  Partial support of the DFG
under contract KU 502/7-1 and the EU Fourth Framework Programme
`Training and Mobility of Researchers', Network `Quantum
Chromodynamics and the Deep Structure of Elementary Particles',
contract FMRX-CT98-0194 (DG 12 - MIHT) is acknowledged.

\end{document}